# The new multimedia educational technologies, used in open and distance learning

**University Lecturer Dieter Penteliuc-Cotoşman**
The West University of Timisoara, Romania,
Faculty of Arts, Department of Multimedia Design

**REZUMAT :** Lucrarea trece în revistă şi se referă la ultimele tehnologii telematice care au transformat Învăţamântul Deschis la Distaţă ş l-au ajutat să devină o alternativă instituţională la cel tradiţional, faţă-în-faţă. Marea majoritate a tehnologiilor, succint prezentate în aici, vor fi implementate în proiectul "ARTeFACt — sistem telematic destinat invatamantului vocational de tip IDD", sistem care va fi lansat oficial la finele anului 2006, in cadrul ofertei instituţionale a Facultatii de Arte a Universităţii de Vest din Timişoara. Coordoarea ştiinţifică a proiectului doctoral «ARTeFACt» aparţine domnului prof.univ.dr.ing. Savii G.George, reprezentând departamentul de Mecatronică al Facultaţii de Mecanică, din cadrul Universităţii "Politehnica" din Timişoara, România.

New Media and the latest IT technologies, the advent of the personal computer, the massive digitization of information and the use of the new telecommunications resources, mainly the Internet, have transformed the world of education.

Traditional schooling does not always suit the needs of the new information society in the context of the present *knowledge-intensive* global economy, where people have to learn and adapt quickly. The increasing thematic diversity and specialization in every domain, as well as the need of the people to learn and work in the same time, also oriented the educators' interest to the use of the computer and to the powerful possibilities offered by the multimedia technologies.

The *individualization* of instruction — a major trend in our days society — recommend the computer, and the computer-related technologies, as the most viable educational tool, capable to support a constructive and active kind of learning. In fact, the multimedia educational technology seems to satisfy all these challenges of the new economical, social and cultural environment in which we live today. Therefore, in spite of some contrary opinions (of those who oppose technology to nature), it is clear now that multimedia offers instruction a great deal of opportunities and fully contributes to a successful learning.

Multimedia technology is a very powerful one. As Tom Boyle shows, "multimedia links the interactive power of the computer to the presentation impact of pictures, sound and motion. It can be delivered on consumer PCs, and multimedia networks beckon us into the future" [Boy97, p. IX].

The computers have become one of the key *instructional technologies*. The various computer applications used by teachers and by students, in formal and informal instruction — from games, simulations, tutorials, word processing and graphic tools, to *integrated learning systems* and *computer based learning environments* — play a multitude of roles and affect in many ways the educational process.

The computer offers some great advantages:
- it provides virtually instantaneous response to student input,
- it can record, analyze and react to student responses,
- it has extensive capacity to store and manipulate information,
- it has the ability to control and integrate a wide variety of media,
- it is able to serve many individual students simultaneously.

These are also some of the principal benefits of learning through new technologies and multimedia systems.

Several studies brought to the conclusion that using multiple means of expression and communication, different sources of information and different teaching methods, is more effective. The combination of multiple and different media and the use of a variety of audio-visual experiences, in correlation with other educational materials, improves considerably the learning process.

Multimedia computers enable educators to built customized and highly interactive learning systems that the students can explore freely, in accordance to their own time, possibilities, interests and criteria. The most important is that students can also manipulate the information within these systems and to adapt the informational material to their learning needs and style.

Computer multimedia *learning systems* are computer-based structured and interactive programs that combine multiple media and integrate them through the IT technology. They incorporate the computer as a display device, management tool, and source of text, pictures, graphics and sound. In these systems, each element complements the others so that the whole is greater than the sum of its parts.

Multimedia learning systems partial simulate the *face-to-face* learning conditions and also create a space-time context filled with a series of multi-sensorial simultaneous experiences that require from the student a multi-sensorial implication.

These systems serve three functions:
- the *presentation* of information,
- the student-teacher *interaction* and
- the *access* to the learning resources.

They are built on the *hypertext* and *hypermedia* technology, which generated a new type of software, called *hypermediaware*. The term *hypertext*, invented by Theodor Nelson in 1974, is used to describe "nonsequential documents composed of text, audio and visual information stored in a computer, with the computer being used to link and annotate related chunks of information (nodes) into larger networks, or webs" [H+96, p. 261].

The hypertext immerses users in a rich information environment, one in which words, sounds and still and motion images can be connected in diverse ways. It allows the development of user centered flexible browsing systems. The *hypertext* generated the *hypermedia*, where the nodes of information may be of any media type.

*Hypermedia* refers to computer software that uses elements of text, graphics, video and audio connected in such a way that the user can easily move within the information.

*Hypermediaware* is software based on the usage of a hypertext environment, which provide the user with possibilities to move within particular set of information without necessary using a predetermined structure or sequence.

Computer hypermedia systems can be used for :
- *browsing* — one can navigate through information and explore features in detail,
- *linking* — users can create their own connections within the information
- *authoring* — users can create their own collections of information, adding or linking text, graphics and audio.

Therefore they suit the personal learning. The hypertext is interactive by its nature, and this interactivity is the source and the essence of its advantages. Hypermedia engages the learner to make choices about moving within the learning material in meaningful ways, it allows users to connect ideas from different media sources and to navigate through information according to their interests and to built their own mental structures based on their exploration. Hypermedia also encourages collaborative work and learning. Hypermedia technologies are very versatile and they can be successfully used in all the areas of the curriculum, in order to achieve different learning goals.

Computer multimedia is based on a series of *off-line* and *on-line* *technologies*, which can support and enhance the learning process:
- the computer-based interactive video (CBIV),
- the compact disc (CD-ROM),
- the compact disc interactive (CDI) technology,
- the digital video interactive (DVI) technology
- the digital video disc (DVD) technology
- the virtual reality (VR)
- the Internet and the World Wide Web.

Every technology — hard and soft, simple and sophisticated media — has its advantages, limitations and range of application, which have to be taken into consideration from an educational perspective and carefully balanced, when media and technology solutions are integrated into instruction.

*Computer-based interactive video (CBIV)* creates a multimedia learning environment in which recorded video material, provided through a videocassette, videodisc or compact disc, is presented under computer control to users who see and hear the pictures and sounds, and also give active responses that affect the presentation. The interactive aspect is provided through computers, various levels of interactivity being available. The learner responds to audio, visual or verbal stimuli through the input devices that include a keyboard, keypad, light pen, bar code reader, mouse, touch-sensitive screen. The interface device provides the link between the computer and the video player, allowing them to communicate.

*Interactive video* is a flexible technology, which combines text, audio, graphics, still pictures and motion pictures in one easy-to-use system that maintains learners' attention, require their participation and engages them in activities. It is also a practical method for individualizing and personalizing instruction. Interactive video is a valuable learning system for tasks that must be shown rather than simply told, and a powerful active resource, especially for the educational situations in which the learner needs to interact with the instruction. Boyle also considers that "video clips can greatly enhance the authenticity of a computer based learning environment", on condition that the learner has the direct control over the video — control "consistent with the learning goals of the context" —, and that the video is "treated as far as possible as a declarative resource", which "is one that can be entered as a number of points and traversed in a number of ways" [Boy97, 178–179].

The *CD-ROM, DVI, CDI and DVD technologies* have many types of applications in the educational process. They are easy to use, compatible with both computer platforms — we refer at Windows and Macintosh OS personal computers —, and capable of storing a large amount of data. DVI, CDI and DVD also have high-quality sound and video. The development of multimedia technology has opened up a lot of possibilities for learning based on virtual experience.

*Virtual reality* is a computer-generated three-dimensional environment where the user — wearing a special headpiece and manipulating a joystick or a special glove — can operate as an active participant. The essence of VR is the expansion of experiences for the user. It provides learners with opportunities to experiment with simulated environments, which are accessible and safe, and to explore places not feasible in the real world.

The educational applications of virtual reality are highly effective, mainly in training physical skills, because in VR, which seems to be the most powerful extension of simulation based systems; the learner is inside the simulation, immersed in a virtual but effective experience. Simulations are excellent for relating the abstract to the concrete, and they can support both exploration and case based learning.

The *Internet* and the *World Wide Web* (WWW) have had a major impact on education. Multimedia online technologies are excellent teaching and learning tools, especially when distance learning is concerned. In fact, online learning (also called *e-learning*, *virtual learning* or *web-based learning*) represents the latest phase in the development of open, flexible and distance learning, which can be defined as an Internet-based teaching and learning system designed for web-based delivery, without face-to-face contact between teacher and learner.

The Internet is an inter-linked, global network of networks that allows computer worldwide to connect to it and to communicate or exchange data, by means of a common protocol: *Transmission Control Protocol — Internet Protocol (TCP/IP)*. The WWW works on the Internet through its own Hypertext Transfer Protocol (*HTTP* or, the secured version, *SHTTP*), and, for most users, it represents the application interface to the Internet. "The WWW acts like a global, distributed hypermedia system" [Boy97, p. 17]. It is an interactive platform that uses various media: text, plain or formatted, hybrid text and graphics documents, color images, still or animated pictures, video, sound, 3-D models, interaction and simulations. The WWW also supports real time, text-based chat and audio and video communication.

The potential of Internet and WWW for education is enormous. These technologies expand, for both learners and educators, the possibilities to access information from an array of sources (databases, libraries, special interest groups), to exchange data, and to communicate with other learners or with experts in a particular field of study, throughout the world. Multimedia technology can be used to support a variety of educational settings, to provide resources and tools for learning.

During the last years, open and distance learning, among many others interdisciplinary activities with a high social impact, progressively implemented the data automatic processing and the organizing of teaching/learning activities through the latest available IT technologies, mainly for *quality assurance* purposes.

In Open&Distance Learning (ODL), the quality assurance concept, regards several of the interactive multimedia learning system's (IMLS) characteristics which are:

- the IML system's capacity to serve and to meet the educational scope's requirements and objectives,
- both parties that are involved in the educational process must know and be aware of the criteria the IMLS uses in order to evaluate the students / teachers perfomances,
- the entire ODL activities are evaluated in certain points, planed in advance and considered to be of a crucial importance, and through IDD particular methods — data gathering, interviews, questionnaire-based statistics, results' analysis of the students performances, while using the ODL interactive system etc. —, in order to support the quality assurance approach.

There is a strong tendency towards ODL software and hardware convergence, especially in the realm of the so-called Telematic Technologies. The huge changes that took place at the end of the end of the XX$^{th}$ century were possible because of two main breakthroughs: on one hand, the digitization, on the other, the appearance of high-speed transfer networks. These two IT technologies discoveries, that had shaped the communication, publishing and entertainment's industries, are, basically, made possible by the computer's capacity to reduce all conventional, analogue, forms of information to a unique logic form, called *streams of digital bits* which are able to represent complex combinations of *data types* (alphanumeric values and text, sounds, dynamic images, simulations, integrated learning and assistance systems etc.). All these technological improvements transformed the personal computer, of a student or of a teacher, into an important *node* belonging to a giant *global network*.

A closer look at the *telematic technologies'* latest developments and tendencies, reveal the fact that they can be classified into two main categories:

- *Hardware developments*, especially those regarding the wireless local networks (WLAN) and their specific data transmission technologies — *IrDA, Bluetooth, DSL, DirectPC, Web Wireless, PowerLine, ISDN, compact video-conferencing systems via IP/ISDN like PictureTell 760 XL-TBR* etc. ;
- *Software developments*, especially those concerning *Server-Side scripting* technologies for web applications, based on dynamic pages, and the realm of *Integrated Teaching/Learning Environment, of e-Learning type* — e.g. *Lotus Learning Space-Virtual Classroom, TopClass e-Learning Suite (WBT)*.

The first category of hardware tendencies had, as a starting point, the incredible opportunities brought by the LAN Wireless type of networks, based on radio waves, as a transmission medium and their propagation in order to support communication and/or accessing the network. The main advantages of WLAN are related to the network administration, the physical placement of the network, the network's bandwidth and its security. By far, the most important tendency, for the ODL type of education, *ISDN*, which stands for *Integrated Services Digital Network*, represents the first step towards a universal digital network that would ensure the integration of data and voice transmission, while using the facilities of a usual phone communications company. For this particular reason, ISDN changes the process of analogue-digital and viceversa conversion that is significantly improved. The old analogue phone system is transformed into a complete digital network, allowing the simultaneous voice and data integrated transmission, from one end to the other of the phone channel. For the individual end-users, potential students of an ODL system, the ISDN technology means many new and useful functions through their actual phone wires: electronic reading devices, video apparatus with slow scanning, Internet access for www browsing, the chance to use a SMTP electronic mail service.

Directly connected with the ISDN technology, by the fact that it is using this kind of integrated digital transmission technology, the *compact video-conferencing systems,* like those produced by the PictureTel Company (e.g. the PT-760 XL-TBR model) have many important characteristics as follows :

- *portability*, because they are self-contained units (the camera and the video codec are integrated in the same base-unit),
- impressive *bandwidth* (up to 384 kbps),
- transmission *speed* (the system is able to transmit video material up to 30 frames/sec. with an ISDN 384 kbps connection),
- in order to see the proximity placed sites, which are participating at the video-conference, the system includes the *"picture-in-picture"* function,
- for a natural and HI-FI quality sound, the system has *superdirective microphones*,
- a mechanic video-camera, with ten default positions, programmed in advance in order to ensure recorded images with balanced detail-general picture framing and composition,
- automatic voice tracking with a video camera which follows the person speaking in a particular moment and the capacity to switch positions as the speaker is changing,
- using the function "freeze frame graphics" for a simultaneous utilisation of another video-camera that is bringing into the conference images from external documents, images from a PC or a video-player,
- the system van be linked to a LAN or WAN through a "10baseT" standard.

If these are the most important latest developments and tendencies in the realm of dedicated hardware for ODL, as we already stated, in the filed of *software* related telematic technologies, those which are to be mentioned regards, among others, *the Web and CD-ROM applications based on server-side scripting* — the so called *dynamic web pages* —, and *Object-Oriented Languages* applied to this kind of applications. Like any other multimedia artefact, these collections of pages/screens/interfaces have everything to do with interaction: with the user/visitor, with other information resources, with similar applications that are resident on the server, with databases. If one is asked to identify the differences between a static, *.html, structuring, tag-based, language page and a dynamic, coded in *php or *.xml, the first aspect that would pop-up in his/hers mind would be about the way the content is predetermined or it is shaped "on-the-fly", as some sort of system's reaction on the user's requests. The applications which are based on dynamic pages are used, in ODL, for three types of functions:

- rapid and simple *information finding* and *discovering* within a content-rich website/application (e.g. www.amazon.com) ;
- data (provided by the users/visitors) recording, storing and gathering,
- web applications that are updating the content of a site who's content is changing on regular basis (e.g. www.economist.com).